# A Possible Resolution of the Tolman Paradox as a Quantum Superposition

Moses Fayngold

*Department of Physics, New Jersey Institute of Technology*

This is an attempt to find a hidden virtue in Tolman's paradox by showing that it can give rise to quantum superposition. We consider tachyon exchange between two particles and show that it can generate superposition of eigenstates characterizing each particle, as well as the entangled state of the particle pair. The new possible aspect of quantum superposition reveals an unexpected connection with cosmological expansion of the Universe.

The problem of superluminal signaling using tachyons as signal carriers has been extensively discussed [1-12]. The general consensus was that the existence of tachyons would violate causality, giving rise to what is known as Tolman's paradox [1-5]. There were also attempts to show that even if such particles do exist, they cannot be used for superluminal signaling [9-15]. In particular, it was shown [15] that causality can be conserved for a charged tachyon if there is a specific relationship between its electric and magnetic charges, similar (but not identical to) Dirac's conditions [16] for a subluminal monopole. However, in this paper we explore a different venue, admitting the possibility of superluminal signaling and exploring its implications. Our analysis uncovers an intriguing possibility to re-interpret Tolman's paradox, which is usually considered as an unambiguous violation of causality, as an underlying mechanism of quantum superposition.

Consider a process which is observed in an inertial reference frame S as a tachyon emission by a stationary object A (event K) and its subsequent absorption by a receding object B (event L). Since the tachyon's world-line between K and L is space-like, in the rest frame S′ of B the temporal order of K and L is reversed under condition

$$V \tilde{v} > c^2 , \qquad (1)$$

where $V$ is the speed of S′ relative to S, and $\tilde{v}$ is the tachyon speed in S. According to the reinterpretation principle [2], the same process is then observed in S' as the tachyon emission from B and its subsequent absorption by A [2, 6], but this does not constitute any paradox for the events connected by a space-like interval [2, 14, 15]. For the sake of simplicity, we henceforth assume that the tachyons' world lines are straight, thus ignoring their acceleration due to gravitational (and, possibly, EM) Cerenkov radiation.

The Tolman paradox arises when we exchange tachyons between two objects in such a way that their world lines form a loop (Fig.1). In the most dramatic example, A emits a primary tachyon $\tau_1$ which hits B, and B responds by emitting its own tachyon $\tau_2$ (secondary tachyon), which hits A and destroys it. Let the response time of B be sufficiently short, and the speed of $\tau_2$ relative to B also satisfy (1). Then the destruction of A precedes (and thereby prevents) the emission of $\tau_1$, creating a paradoxical situation: if A emitted $\tau_1$, then A must have been destroyed earlier and could not have emitted it; on the other hand, if A did not emit $\tau_1$ due to earlier destruction, then B never responded, so A was not destroyed and must emit $\tau_1$. Thus A is suspended in limbo, and one cannot tell whether or not A exists.

If we take this contradiction as an argument against the existence of tachyons, the paradox is successfully avoided. But here we propose another logical possibility by pointing out that the situation where a particle's existence is indeterminate is well known in Quantum Mechanics. In Schrodinger's famous thought experiment (the Schrodinger cat [17]), the cat's life is similarly suspended in limbo (superposition of two states – being dead and being alive) until a measurement determines its state. This similarity suggests the idea of describing the superposition of particle states in terms of tachyon exchange.

Our approach has so far not found any indications that tachyons might provide the missing superluminal mechanism which, according to Bell's theorem [18], would be necessary in any attempts to describe Quantum Mechanics in terms of hidden variables [19, 20]. Nevertheless, the identification of tachyons as the superluminal signaling tool responsible for quantum superposition could be a tempting idea for several reasons.

First, the conventional treatment of superposition of eigenstates, including the macroscopic cases like the Schrodinger cat would be at least as self-intuitive as the particle caught in a time loop.

Second, tachyons could offer another view of entanglement, opening the way for interpreting particles A and B in Tolman's paradox as an entangled EPR particle pair.

Third, the concept of quantum measurement with the accompanying problem of an instantaneous collapse of the wave function would admit a more intuitive interpretation as the absorption or deflection of a tachyon, which would disrupt the time loop.

And finally, the new perspective, unless shown a blind alley, may offer new insights into Relativistic Quantum Field Theory, Quantum Information, and other research areas.

In what follows, we treat tachyons without making any preliminary assumptions about their quantum-mechanical properties. This leaves open the question whether this upstart view of tachyons as relativistic *classical* particles is in fact permissible.[1]

We will now focus on the notion of quantum superposition as a *resolution* of the Tolman paradox. In view of the above-mentioned analogy with the Shrodinger cat, we first personify A as the Cat and B as the Hunter (Fig. 2).

The Cat is stationary in S, whereas the Hunter is receding at a speed $V$. The Cat radiates a tachyon $\tau_1$ at a moment $t = t_K$ (event K). The tachyon reaches the Hunter at some later moment $t = t_L$ (event L). This alerts the Hunter, and he shoots back tachyon $\tau_2$ at a moment $t = t_M \geq t_L$ (event M). Tachyon $\tau_2$ is so fast in S' that its world line in S is represented by segment MN with N preceding K. The tachyon hits and kills the Cat at $t_N < t_K$ (event N). In frame S, this is observed as a spontaneous emission of a tachyon from N to M, resulting in Cat's death. Tolman's paradox thus arises because if the Cat had been killed before $t = t_K$, he cannot emit $\tau_1$, the Hunter remains idle and does not kill the Cat. If the Cat is not killed, it emits $\tau_1$, alerts the Hunter and gets killed. We use the term "limbo" to refer to the paradoxical state of the Cat (and the Hunter and both tachyons) in the described time loop. If superluminal signaling does exist, then the only way to eliminate the paradox is to *introduce* the concept of quantum superposition. This immediately converts the supposedly classical system (the Cat, the Hunter, and two tachyons) into quantum-mechanical one. The whole loop can then be described as a *superposition of eigenstates*. We introduce the following notations:

$$\left.\begin{aligned}&|C_{alive}\rangle = \text{alive cat};\ |C_{dead}\rangle = \text{dead cat};\ |H_{idle}\rangle = \text{idle hunter};\ |H_{alert}\rangle = \text{alert hunter};\\&|C_{limbo}\rangle = \text{cat in a limbo state};\ |H_{limbo}\rangle = \text{hunter in a limbo state};\\&|\tau\rangle = \text{emitted tachyon};\ |0\rangle = \text{no tachyon};\ |\Psi\rangle - \text{the state function of the whole system}\end{aligned}\right\} \quad (2)$$

---

[1] The known models of tachyons based on Quantum Field Theory impose important limitations on their properties. Thus, a self-consistent model by G. Feinberg [4] describes tachyons as spinless particles obeying the Fermi-Dirac statistics. However, since Feinberg employs the Klein-Gordon equation for tachyons as the underlying assumption, the model cannot be taken as a definitive proof that tachyons must have zero spin. By the same token, the condition $|p| \geq m_0 c$ for the tachyon momentum would preclude tachyons from being local particles only if we assume in advance that tachyons must be described as a superposition of their de Broglie waves.

As seen from Fig. 2, the temporal dimension breaks up into 5 distinct regions with different expressions for $|\Psi\rangle$. In frame S, assuming the superpositions to be equally-weighted, dropping the normalizing factor and restricting the phase shift $\alpha$ between the superposed states to $\alpha = 0$ or $\pi$, we have:

$$|\Psi(t < t_N)\rangle = |C_{alive}\rangle |H_{idle}\rangle |0_1\rangle |0_2\rangle, \quad \text{(Region I)} \quad (3.\text{I})$$

$$\begin{aligned}|\Psi(t_N < t < t_K)\rangle &= |C_{alive}\rangle |H_{idle}\rangle |0_1\rangle |0_2\rangle \pm |C_{dead}\rangle |H_{idle}\rangle |0_1\rangle |\tau_2\rangle = \\ &= (|C_{alive}\rangle |0_2\rangle \pm |C_{dead}\rangle |\tau_2\rangle) |H_{idle}\rangle |0_1\rangle,\end{aligned} \quad \text{(Region II)} \quad (3.\text{II})$$

$$\begin{aligned}|\Psi(t_K < t < t_L)\rangle &= |C_{alive}\rangle |H_{idle}\rangle |\tau_1\rangle |0_2\rangle \pm |C_{dead}\rangle |H_{idle}\rangle |0_1\rangle |\tau_2\rangle = \\ &= (|C_{alive}\rangle |\tau_1\rangle |0_2\rangle \pm |C_{dead}\rangle |0_1\rangle |\tau_2\rangle) |H_{idle}\rangle,\end{aligned} \quad \text{(Region III)} \quad (3.\text{III})$$

$$\begin{aligned}|\Psi(t_L < t < t_M)\rangle &= |C_{alive}\rangle |H_{alert}\rangle |0_1\rangle |0_2\rangle \pm |C_{dead}\rangle |H_{idle}\rangle |0_1\rangle |\tau_2\rangle = \\ &= (|C_{alive}\rangle |H_{alert}\rangle |0_2\rangle \pm |C_{dead}\rangle |H_{idle}\rangle |\tau_2\rangle) |0_1\rangle,\end{aligned} \quad \text{(Region IV)} \quad (3.\text{IV})$$

$$\begin{aligned}|\Psi(t > t_M)\rangle &= |C_{alive}\rangle |H_{alert}\rangle |0_1\rangle |0_2\rangle \pm |C_{dead}\rangle |H_{idle}\rangle |0_1\rangle |0_2\rangle = \\ &= (|C_{alive}\rangle |H_{alert}\rangle \pm |C_{dead}\rangle |H_{idle}\rangle) |0_1\rangle |0_2\rangle\end{aligned} \quad \text{(Region V)} \quad (3.\text{V})$$

Initially (Eq. (3.I)), the Cat is alive, and the Hunter is idle. During the time interval $t_N < t < t_K$ between events N and K (Eq. (3.II)), the Cat and $\tau_2$ are suspended in limbo, forming an entangled superposition, the tachyon $\tau_1$ has not yet been emitted, and the Hunter has yet to be alerted. At $t_K < t < t_L$ (Eq. (3.III)), the Hunter is still idle, and the Cat is suspended between life and death as the previous state. The difference from the previous time interval is that now we have an entangled superposition involving the Cat and *two* tachyons: in each respective basis state of the whole system, if one tachyon exists, the other does not, and vice versa. If our devices are not tuned to detect the tachyons, then there is no observational difference between regions II and III, and in both of them we observe the cat in pure superposition.

During $t_L < t < t_M$ (Eq. (3.IV)) tachyon $\tau_1$ no longer exists, $\tau_2$ is in limbo, and we have an entangled superposition between the Cat, the Hunter, and $\tau_2$. Finally, after $t = t_M$ (Eq. (3.V)), the Cat winds up in an entangled superposition with the Hunter.

In the considered cases the state function is not symmetric with respect to the Cat and the Hunter. But this only reflects the asymmetry in the initial conditions which mandate the Cat to be killed in the $C \leftrightarrow \tau_2$ interaction, whereas the Hunter survives interaction with both tachyons.

Summarizing the presented argument, we can say that some cases of quantum superposition might, at least in principle, originate from tachyon exchange between two objects.

The results for the exotic case of the Schrodinger cat (and the Hunter) can be extended onto some real particles A and B. One (and most natural) extension would be getting back to a radioactive particle A whose decay (or absence of thereof) in [17] gives rise to the dubious state of the Schrodinger cat. The above analysis shows that quantum superposition of the decayed and non-decayed states of A (when not observed) could originate from a tachyon exchange with another particle B. Its description would be given by the same system of Equations (3.I-V) if we use the similar assumptions about particle interactions with the tachyons: absorption of $\tau_1$ transfers B from a lower-energy state $B_1$ to a higher-energy state $B_2$, which enables it to emit $\tau_2$ with an appropriate energy; absorption of $\tau_2$ by A is recorded as its emission in S, and this emission is totally spontaneous with respect to the preceding events on the world line of A. It can therefore be considered as an indication of A's spontaneous decay at $t = t_N$, making it impossible for A to emit $\tau_1$ at a later moment $t_k$. Then Eq-s (3.I-V) will describe a superposition of the corresponding quantum states of A and B if we replace:

$$|C_{alive}\rangle \to |A_{non-decayed}\rangle; \quad |H_{idle}\rangle \to |B_1\rangle; \quad |C_{dead}\rangle \to |A_{decayed}\rangle; \quad |H_{alert}\rangle \to |B_2\rangle \qquad (4)$$

Alternatively, each particle can be an atom undergoing transition between the lower and higher energy state via tachyon exchange rather than photon exchange. In this case, a proper choice of the initial conditions would lead to the same situation as considered above, for instance, atom A in a superposition of two distinct states $A_1$ and $A_2$ with energies $E_1$ and $E_2$, respectively – a purely quantum-mechanical effect.

As mentioned, we have considered the special case in which a particle has equal probability $P_1 = P_2 = 1/2$ of being in either of the two distinct states. This, together with the above-mentioned restriction on phases, determines the corresponding probability amplitudes $c_1 = \pm c_2 = 1/\sqrt{2}$ in the superposition. A general superposition $|\Psi_A\rangle = c_1|A_1\rangle + c_2|A_2\rangle$ with arbitrary (normalized) complex amplitudes, may require to describe each tachyon as a wave, just as we do in perturbative theory of atom interaction with light described as a EM wave (but the velocity of a signal carried by the tachyon waves must be superluminal).[2] This step may modify the amplitudes, since the tachyon $\tau_2$ can now be considered as a part of "primary" wave $\tau_1$ back-scattered from atom B with complex scattering amplitude including the phase shift $0 \leq \alpha \leq \pi$. The specific value of $\alpha$ will be determined by the form of interaction between $\tau_1$ and B. There will also appear time indeterminacy in the interaction events K and N (as well as L, M). This would be consistent with quantum-mechanical properties ascribed to the tachyon the moment we introduce the possibility of superposition.

Alternatively, we can suppose that a tachyon gets absorbed and reemitted with some probability which will also modify the resulting probability amplitudes of the superposed states.

Yet another possibility is to introduce additional time loops associated with several primary tachyons emitted at different moments of time. In the first example, after $t = t_N$ we only have a 50% chance to find the Cat alive. Thus a second loop initiated at a later time will already involve a smaller probability amplitude to find the Cat alive. By reiterating this process, we can modulate probability amplitude of the Cat being alive as a function of time.

Generally, the suggested idea may have many ramifications. We do not see immediate answers to many questions, for instance, whether it is possible to describe consistently quantum-mechanical superposition of more than two states in terms of tachyon-tardyon interactions involving more than two tachyons.

It is not clear how if at all, the suggested mechanism could produce, for instance, a specific superposition of 2 eigenstates of a linearly polarized photon to give rise to a state with a circular polarization. The similar question arises for electron spin states. In such cases the tachyon exchange with environment might reveal the underlying mechanism for incompatibility of different spin components. This, however, might require the existence of tachyons with non-zero spin.

But some of the questions can be answered already at this stage. For instance, we now have a convenient way to visualize the process of measurement. If we put a tachyon detector between the two particles, then as soon as one of the tachyons gets absorbed in it, the time loop is destroyed and the other tachyon never materializes. Each particle then collapses into one of the two possible states, depending on which tachyon gets absorbed in the detector. Alternatively, it is conceivable that nature itself could take care of the process of measurement if a tachyon gets scattered or absorbed in the way, thus never arriving to its destination [12, 13] and thereby failing to close the time loop. Some of these questions will be addressed in a separate article.

An important general feature of the described process is that it can only work for particles A and B receding from one another. Therefore, in order for the universal phenomenon of quantum superposition to be manifestation of the Tolman paradox at work, for each A there has to be an abundance of receding particles with an appropriate energy spectrum. Such abundance is provided by cosmological expansion of the Universe. If this expansion were not known today, it could be predicted by the suggested scheme as a necessary condition for it to work on a universal scale. This raises an additional question of whether and how the known time arrow within standard Quantum Mechanics itself [22 -24] could be related to the cosmological arrow of time.

The possibility of superluminal signaling remains a sensitive and sometimes even controversial issue [25 - 31]. The observation pointed out in this article may add another dimension to the wide spectrum of discussions including superluminal tunneling [32- 35], quantum entanglement [36 - 38], direct action [39], and other topics.

---

[2] Once we have admitted possibility of superluminal signaling, we must leave out, at least temporarily, the question of how it can be consistent with the known restriction on the propagation speed of a shape discontinuity in signal-carrying waves [21], or, possibly, refine the definition of a signal.

To summarize, we see that the conventional view of Tolman's paradox as the decisive argument against the possibility of superluminal signaling can as well be reinterpreted as the underlying mechanism of quantum superposition. In this paradigm, what was deemed a contradiction emerges instead as the basic building block of the quantum theory, pointing once again to the relativistic underpinnings of Quantum Mechanics. We are driven to conclude that superluminal particles could conceivably lie at the basis of some purely quantum-mechanical effects. In this respect, Quantum Mechanics could be considered as Nature's device against any violations of relativistic causality.

**Acknowledgements**
I wish to thank David Green (The New School University) for valuable discussions and support of this work.

**Figures**

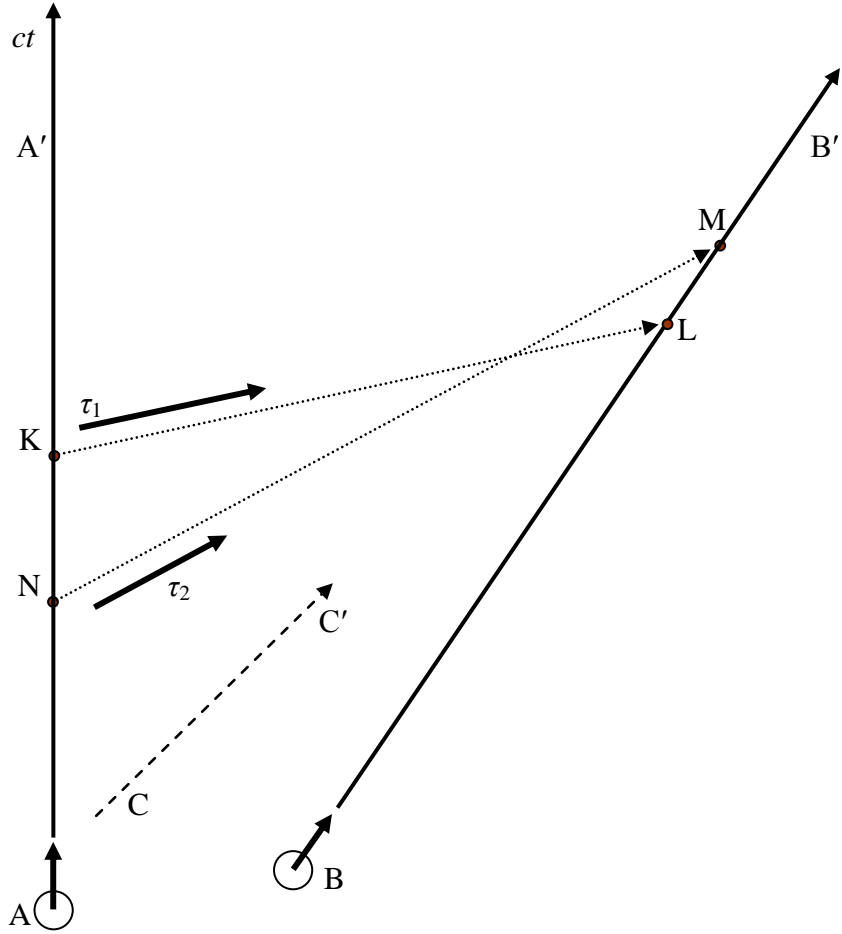

**Fig. 1**
Objects A and B caught in a time loop KLMN.
AA′ and BB′ are the world-lines of A (frame S) and B (frame S′), respectively. KL is the world-line of the primary tachyon $\tau_1$ emitted by A. NM is the world-line of the secondary tachyon $\tau_2$ emitted by B. Since it is emitted back at A, this world-line intersects with *ct*-axis and accordingly, under condition (1), hits A *before* the emission of the primary tachyon. In frame S, this is observed as *spontaneous* emission of tachyon $\tau_2$ towards B, which precludes the emission of $\tau_1$. The directions of both tachyons *as observed in S* are shown by arrows. The recoil effects are assumed to be negligible.
CC′ is the world line of a photon.

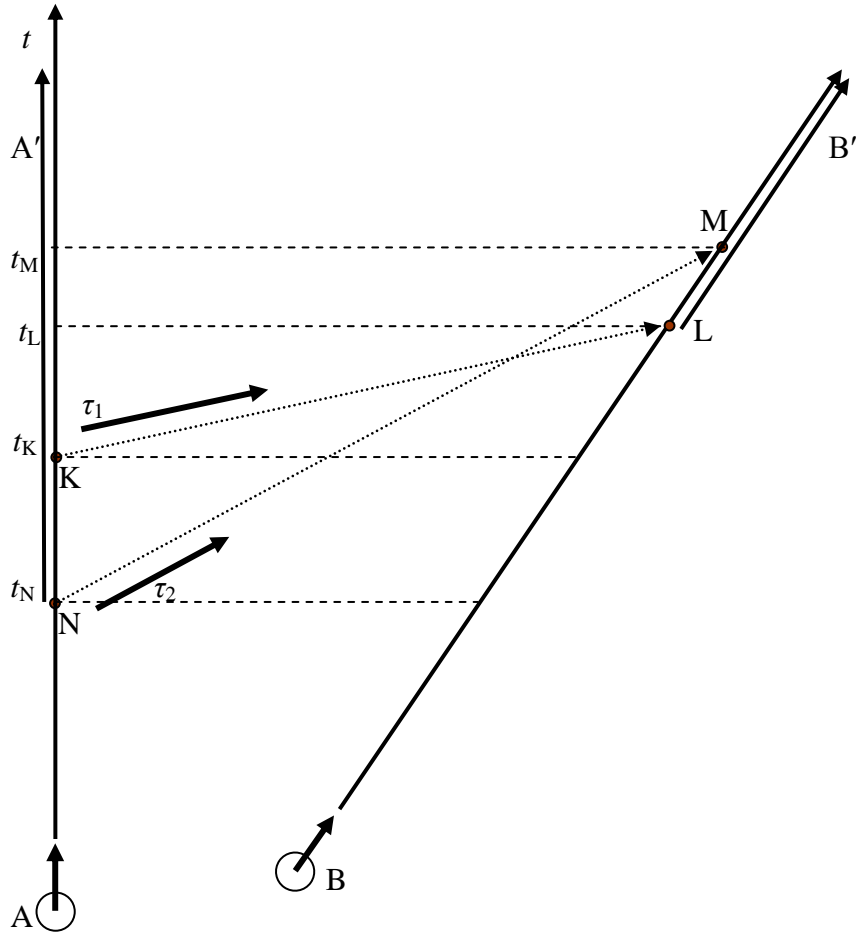

**Fig. 2**
The same as Fig. 1, but explicitly showing time domains associated with events K, L, M, N. The temporal axis of frame S is denoted here just as $t$. Also shown are the world lines of the additional basis states $|C_{dead}\rangle$ and $|H_{alert}\rangle$ emerging from events N and L, respectively, due to tachyon exchange between A and B.